# ZEUS
# A Domain-Oriented Fact Comparison Based Authentication Protocol

**Document version:** 1.0
**Document author:** Kirti Chawla
**Document Signature:** 0915063205

# Exploring a domain-oriented facts comparison as means of zero-knowledge protocol

In this paper, facts existing in different domains are explored, which are comparable by their end result. Properties of various domains and the facts that are part of such a unit are also presented, examples of comparison and methods of usage as means of zero-knowledge protocols are given, finally a zero-knowledge protocol based on afore-mentioned concept is given.

## Domains

A domain is aggregation of knowledge body. It may belong to a person or available in the form of electronic or non-electronic based stored information. A person can be a knowledge body, so can be a book and a computer disk. Let D as a set be domain, which has following properties.

1. It has infinite members
2. It is characterized by a defining property that holds true for all members of set
3. Members are closed under respective domains

Notation-wise, a series of domains is given by, $[D_i]$, where [i] is type of a domain. A domain may define a set of operators that are applicable on its member. Domains can be infinite. A property is defined by {Property}.

## Facts

A fact is member of domain. Alternatively, a domain is collection of facts. A collection of facts forms a knowledge body. Let F as an element of domain, be Fact, which has following properties.

1. It belongs to a domain D
2. It shares certain properties with other facts in a domain
3. Facts when operated upon by a defined operator, it results into facts that are closed under pertinent domain

Notation-wise, an infinite series of facts is given by $[F_i]$, where [i] is instance of fact.

## Domain-Oriented Fact Comparison

Two domains $[D_i]$ and $[D_j]$ may be same or different, but comparable on the conditions, which are given as under:

1. The facts of domain [i] and [j] at instance [k] in both have same outcome, irrespective of domain. Here outcome is concerned with only magnitude of values.
2. The operation in one domain has mapping in terms of magnitude of value to another domain. This can be mapping of multiple facts from one domain onto single facts in another domain.

## Example

Let $[D_i]$ be {Markets} and $[D_j]$ be {Space exploration}. $[D_i]$ and $[Dj]$ have following members:

$[D_i]$ = ("21$^{st}$ Avenue Road", "Coles Park", "M.G Road" …)
$[D_j]$ = ("Mars Pathfinder", "Cassini-Huygens", "Nebula" …)

Facts in $[D_i]$ are 21$^{st}$ Avenue Road, Coles Park and so forth. Facts in $[D_j]$ are Mars Pathfinder, Cassini-Huygens and so forth. The domains which are similar will have similar set of facts.

## Zeus: A zero-knowledge protocol using Domain-Oriented Fact Comparison

Let's assume a hypothetical conversion between two persons Tom and Mary. The thread of conversation is as follows:

Tom: Where did you go today for shopping ?
Mary: The same place you go for boozing

Tom: Moon is waning
Mary: Bring me a cold-cream

Mary: Are you two-timing me ?
Tom: Are you going to church ?

Mary: I have brought something for you
Tom: Is that same as yesterday ?

…

As indicative of the conversation thread, both parties exchange information in a series of question-response, response-response, question-question and response-question. The things evident from the conversation are as under:

1. There is no explicit exchange of information
2. Both parties get the answers implicitly
3. There is an implied conversion from a fact of a domain to a fact in another domain

*Zeus* is a small protocol, which takes advantage of the underlying phenomenon of domain-oriented fact comparison to test the party of possessing knowledge, which interests the questioner. First, protocol basics are formalized, then protocol is introduced and finally it is tested with scenario.

# Algorithm

Let's introduce some preliminaries, before the algorithm is presented. They are given as under:

1. *User(s)*: The communicating parties are called Users. In a 2-party conversation there is only one (1) instance of conversation, whereas a given N-party conversation can be reduced to M $\{1 < M < {}^NC_2\}$ instances of 2-party conversations. Two different users are defined as $U_i$ and $U_j$.

2. *Fact(s)*: A basic unit of data-exchange between two (2) users $\{U_i$ and $U_j\}$. There two types of facts {Fact-In-Question-Form, Fact-In-Answer-Form}. They are defined as $F_1$ or $F_2$ depending whether the fact is question or answer.

3. *Confidence-measure*: The confidence measurement is metric used against a given threshold and guarantees valid authentication to legitimate users.

4. *Mutually Known or Unknown Domain*: A universal set of facts known or unknown to parties engaged in conversation or intending conversation.

5. *Rendezvous*: A state which defines a conversation happening between intended parties.

Let's now give a workable scenario that utilizes the afore-mentioned entities engaged in a conversation. First a real-life example is given then the preliminaries are mapped to pertinent place-holders.

"*Tom* and *Mary* are two persons intending to start a *conversation*. Mary is furious, because she feels Tom didn't kept the promise of taking her to movie previous day. Tom intends to break this barrier because he wants to open a joint account with her in a bank, but for her to listen to all the talk, she has to be in good mood. Currently, Mary has low *confidence* in Tom." With this precedent, the conversation starts.

Tom: How are you feeling now ?
Mary: Not the same as you are

Tom: How can you be sure about my state-of-art ?
Mary: Because it's not the same as mine.

Tom: I had to go to a place you visited last week for some important work.
Mary: Really ! I thought you were in the same drab office.

Tom: I feel the same as that day when you were in similitude state as today
Mary: And, I think I feel the same towards you as I did that day

Tom: Can we open a bank account together ?
Mary: Would you like to have some *Naïve-la-presto* for dinner ?

...
At the end of conversation Mary agrees with Tom for opening a joint-account in a bank. So, Tom gradually increases Mary confidence in him and hence authenticates himself to prove genuine-ness.

In a way, one can replace *italics* place-holder with pertinent preliminaries. Let Tom and Mary be *Users*. Let whole conversation be a *rendezvous* point. The conversation between them is full of *facts* in either form {Facts-In-Question-Form, Facts-In-Answer-Form}. During the conversation, the *confidence* of intended parties increase and this leads to mutual agreement and hence proves the *genuine-ness* of one party to another. All this time, they are exchanging facts based on a *Mutually Known or Unknown Domain* called "Tom's inability to take Mary out for a movie on a previous day".

The state-of-art now enables, to put forth the algorithm in its true form. It is given as under:

Step 1: $U_i$ initiates conversation with $U_j$

Step 2: Let the conversation continue and correlate the facts from users
- SET Authentication ← NOT_DONE
- While (($C_{ip}$ Less-Than-Equals I_Threshold) and ($C_{jp}$ Less-Than-Equals J_Threshold))
- Begin
  - $U_i$ sends $F_{[1|2]k}$ to $U_j$, which is chosen from $D_i$
  - $U_j$ sends $F_{[1|2]m}$ to $U_i$, which is chosen from $D_j$
  - $F_p \leftarrow F_{[1|2]k} \ominus F_{[1|2]m}$
  - If ($F_p$ in T) Then
  - Begin
    - $C_{ip} \leftarrow C_{ip} + 1$
    - $C_{jp} \leftarrow C_{jp} + 1$
    - $D_i \leftarrow D_i \cup \{F_p\}$
    - $D_j \leftarrow D_j \cup \{F_p\}$
  - End
  - Else
    - Continue
- End

Step 3: Check if genuine-ness is proved and stop
- If (($C_{ip}$ Greater-Than-Equals I_Threshold) and ($C_{jp}$ Greater-Than-Equals J_Threshold)) Then
- Begin
  - SET Authentication ← DONE
- End
- Stop

The notations are described as under:
U: User
F: Fact { $F_p$: Resultant Fact }
C: Confidence Measure
D: Mutually Known or Unknown Domain
T: Partially Known Set of Target Facts
X_Threshold: Individual Threshold